\documentclass[10pt,conference, screen]{IEEEtran}

\usepackage{amsmath,amsfonts,bm}

\def\eqref#1{equation~\ref{#1}}
\def\1{\bm{1}}

\DeclareMathAlphabet{\mathsfit}{\encodingdefault}{\sfdefault}{m}{sl}
\SetMathAlphabet{\mathsfit}{bold}{\encodingdefault}{\sfdefault}{bx}{n}

\usepackage{algorithm}
\usepackage{hyperref}
\usepackage{bigstrut}
\usepackage{subcaption}
\usepackage{url}
\usepackage{cancel}
\usepackage{algpseudocode}
\usepackage{amsmath,amsfonts,amsthm,epsfig,epstopdf,url,array}

\usepackage{listings}
\usepackage{xcolor}
\usepackage{enumitem}

\newcommand{\ic}[1]{\begin{small}\texttt{#1}\end{small}}
\definecolor{codegreen}{rgb}{0,0.6,0}
\definecolor{codegray}{rgb}{0.5,0.5,0.5}
\definecolor{codepurple}{rgb}{0.58,0,0.82}
\definecolor{backcolour}{RGB}{242,242,242}

\lstdefinestyle{mystyle}{
    language=Python, 
    backgroundcolor=\color{backcolour},   
    commentstyle=\bfseries\color{codegreen},
    keywordstyle=\bfseries\color{magenta},
    numberstyle=\footnotesize\color{codegray},
    stringstyle=\bfseries\color{codepurple},
    emph={l, sorted, x},
    emphstyle=\color{red},
    basicstyle=\ttfamily\footnotesize,
    breakatwhitespace=false,         
    breaklines=true,                 
    captionpos=b,                    
    keepspaces=true,                 
    numbers=left,                    
    numbersep=5pt,
    xleftmargin=10pt,
    showspaces=false,                
    showstringspaces=false,
    showtabs=false,                  
    tabsize=2,
    frame=lines,
    morekeywords={assert, True, False},
}

\lstdefinestyle{mystyle-simple}{
    language=Python, 
    backgroundcolor=\color{backcolour},   
    commentstyle=\bfseries\color{codegreen},
    keywordstyle=\bfseries\color{magenta},
    stringstyle=\bfseries\color{codepurple},
    emph={sort},
    emphstyle=\color{red},
    basicstyle=\ttfamily\footnotesize,
    breakatwhitespace=false,         
    breaklines=true,                 
    captionpos=b,                    
    keepspaces=true,
    showspaces=false,                
    showstringspaces=false,
    showtabs=false,                  
    tabsize=2,
    frame=lines,
    morekeywords={assert, True, False},
}

\theoremstyle{plain}

\theoremstyle{definition}
\newtheorem{defn}{Definition}[section]

\theoremstyle{remark}

\begin{document}

\title{Interactive Code Generation via Test-Driven User-Intent Formalization}
\DeclareRobustCommand*{\IEEEauthorrefmark}[1]{%
  \raisebox{0pt}[0pt][0pt]{\textsuperscript{\footnotesize #1}}%
}

\author{
    \IEEEauthorblockN{
    Shuvendu K. Lahiri\IEEEauthorrefmark{1}\textsuperscript{\textdagger}, 
    Sarah Fakhoury\IEEEauthorrefmark{1}\textsuperscript{\textdagger},
    Aaditya Naik\IEEEauthorrefmark{2}\textsuperscript{\textsection}, 
    Georgios Sakkas\IEEEauthorrefmark{3}\textsuperscript{\textsection} 
    \\ Saikat Chakraborty\IEEEauthorrefmark{1}, 
    Madanlal Musuvathi\IEEEauthorrefmark{1}, 
    Jeevana Priya Inala\IEEEauthorrefmark{1}, 
    Piali Choudhury\IEEEauthorrefmark{1} 
    \\ Curtis von Veh\IEEEauthorrefmark{1}, 
    Chenglong Wang\IEEEauthorrefmark{1}, 
    Jianfeng Gao\IEEEauthorrefmark{1}}
    \IEEEauthorblockA{\IEEEauthorrefmark{1}Microsoft Research
    \\\{shuvendu, sfakhoury, saikatc, madanm,  jinala, pialic, curtisvv, chenwang, jfgao\}@microsoft.com}
    \IEEEauthorblockA{\IEEEauthorrefmark{2}University of Pennsylvania
    \\ asnaik@seas.upenn.edu}
     \IEEEauthorblockA{\IEEEauthorrefmark{3}University of California, San Diego
    \\ gsakkas@eng.ucsd.edu}
}

\newcommand{\fix}{\marginpar{FIX}}
\newcommand{\new}{\marginpar{NEW}}

\newcommand{\shuvendu}[1]{{\color{red} SKL: #1}}
\newcommand{\saikat}[1]{{\color{brown} GS: #1}}
\newcommand{\sarah}[1]{{\color{blue} SF: #1}}
\newcommand{\TOOL}{\textsc{TiCoder}}
\newcommand{\besttool}{\TOOL}
\newcommand{\neuraltool}{\textsc{NeuralTiCoder}}
\newcommand{\tappProblem}{{\sc ITDCG}}
\newcommand{\pass}[2]{\texttt{pass@#1@#2}}
\newcommand{\passnk}{\pass{k}{m}}
\newcommand{\passdefault}{\pass{1}{1}}
\newcommand{\passk}[1]{\texttt{pass@#1}}
\renewcommand{\algorithmicrequire}{\textbf{Input:}}
\renewcommand{\algorithmicensure}{\textbf{Output:}}

\maketitle
\thispagestyle{plain}
\pagestyle{plain}
\begingroup\renewcommand\thefootnote{\textdagger}
\footnotetext{Equal Contribution.}
\endgroup

\begingroup\renewcommand\thefootnote{\textsection}
\footnotetext{Work done while interning at Microsoft.}
\endgroup

%%%%%%%%%%%%%%%%%%%%%%%%%%%%%%%%%%%%
\begin{abstract}
Large language models (LLMs) have shown great potential in automating significant aspects of coding by producing natural code from informal natural language (NL) intent. However, when interacting with LLMs, users have no guarantees that the code suggestions produced correctly satisfy the intent they provided.   
In fact, it is hard to define a notion of correctness since natural language can be ambiguous and lacks a formal semantics.

In this paper, we propose the workflow of {\it interactive test-driven  code generation} (\tappProblem{}), which leverages lightweight user feedback to (a) formalize the user intent using generated tests that can be useful for debugging, and  (b) produce an improved set of code suggestions by pruning and ranking candidate code suggestions. 

We describe a language-agnostic abstract algorithm for \tappProblem{}, and a concrete implementation \TOOL{}. 
We perform an automated evaluation of \TOOL{} on the \emph{MBPP} and \emph{HumanEval} code generation benchmarks.

Our results are promising with using the OpenAI Codex LLM: our best algorithm improves the \passk{1} code generation accuracy (in absolute percentages) between $22.49\%$ to $37.71\%$ for MBPP and between $24.79\%$ to $53.98\%$ for HumanEval using between 1 to 5 simulated user queries. 
\end{abstract}

\section{Introduction}
\label{sec:intro}

\emph{Large Language Models} (LLMs) have shown tremendous potential in generating natural-looking programs from informal intent expressed in natural language. 
There has been surge in research on training large language models over programming language artifacts in just the last couple of years~\cite{codex_2021, palm_2022, codegen_2022, incoder_2022, polycoder_2022}. 
%Most of these LLMs are based on recent advances in Transformer neural network architectures~\cite{vaswani_2017} and the availability of large corpus of source code in open source (say, GitHub). 
Commercial offerings such as Copilot~\cite{copilot_2022} are now available, and are already known to generate non-trivial fraction of code in real-world developer scenarios~\cite{copilot_maps_2022}. 
% Google paper == palm_2022?

However, the rise of code synthesis from natural language poses new challenges for generating correct code. 
First, natural language is ambiguous, unlike formal specifications, to express the \emph{user intent}. 
Consider the following docstring, taken from MBPP\cite{austin2021program}, a popular Python programming tasks benchmark: 
\begin{lstlisting}[style=mystyle]
def text_lowercase_underscore(text):
    """Write a function that returns true if the input string contains sequences of lowercase letters joined with an underscore and false otherwise""
\end{lstlisting}
%\texttt{"Write a function that returns true if the input string contains sequences of lowercase letters joined with an underscore and false otherwise"}.  
This docstring is inherently ambiguous as 1) it does not specify how many sequences of lowercase letters joined by underscores are expected, and 2) if the input string must consist entirely of the sequence described, or can be a substring.  An LLM may interpret the docstring in a number of ways, and generate several plausible programs with different behaviours.
%As a simple example, the Python docstring \texttt{"""Sort a list of integers"""} does not specify if the user wishes to sort the list in the ascending or descending order of the values. 
More importantly, the lack of a precise semantics of natural language means that one cannot even articulate the correctness of the code generated by a LLM. 
Research has shown that users struggle to understand LLM generated code suggestions presented to them, unless relying on the ability to run or debug the code to internalize program behaviour~\cite{tianyi_2022}. Otherwise, users may accept buggy code, or reject correct code that are too difficult to understand.
Finally, when presented with a long list of plausible suggestions from LLMs, often sorted by some {\it naturalness} of the answer~\cite{hindle2016naturalness}, a user often has to  linearly scan each code suggestion, identify, and reject the incorrect ones. 

While these issues are nascent to the LLM-based code generation space, the problem of disambiguation of user intent has been a long standing challenge in Programming by Example (PBE) paradigm~\cite{Gulwani-pbe-16}, where users provide a set of input-output examples to express intended program behavior. However, prior research has shown that it can be difficult for users to manually provide a sufficient number of examples\cite{lau2009programming} and for those examples to sufficiently cover the hypothetical input space\cite{lee2017towards}. An {\it augmented} set of examples (or inputs) that cover a wider range of the input space can not only help to better guide synthesis techniques by pruning away more plausible programs, but also helps to reduce the mental load of understanding and validating synthesized programs\cite{zhang2020interactive} \cite{le2017interactive}.

Although the benefits of intent disambiguation through example generation has been studied in the context of specific DSLs in PBE (such as regular expressions), such techniques do not readily apply to the problem of code generation in arbitrary programming languages from natural language intent. 
First, there is a lack of formal and symbolic reasoning engine for mainstream programming languages that allows solving for generating distinguishing/disambiguating examples.
Second, users typically do not  augment natural language intent with tests or examples to restrict the space of solutions~\cite{perry2022users}.
%and (b) \sarah{current developer-AI interaction models do not explicitly support the scenario where} users can provide an initial set of examples to restrict the space of solutions\shuvendu{Can we get some reference that users seldom mix NL with tests when using Copilot like tools?} 
% while previosu techniques of PS are domain specific LLMs open this to a wide variety of programs
%\shuvendu{Motivation from prior works motivation}

Inspired by these findings around example generation and disambiguation techniques in PBE~\cite{zhang2020interactive}, and recent emerging ability of LLMs to generate tests~\cite{lemieux2023codamosa,dinella2022toga,schafer2023adaptive} \textbf{in this paper, we advocate \emph{leveraging lightweight user-feedback} through tests (as a form of weak formal specification) to improve correctness of LLM-generated code}. 
Specifically, we advocate the workflow of {\it interactive test-driven code generation} (\tappProblem{}) to  create a framework to (a)  formalize user intent through generated tests, and also (b) generate a {\it ranked} list of code that is consistent with such tests. 

% \begin{figure*}[t]
%     \centering
%     \includegraphics[width=0.9\linewidth]{intro-example.png}
%     \caption{\sarah{placeholder for now. Either as listing in text or one figure.}}
%     \label{fig:example}
% \end{figure*}

%\shuvendu{Find a better example} 
Let us demonstrate a simple instantiation of the intuitive framework using the earlier example, taken from MBPP\cite{austin2021program}, where a user prompts an LLM to generate code satisfying their natural language intent.
%\begin{lstlisting}[style=mystyle]
%def text_lowercase_underscore(text):
%    """Write a function that returns true if the input string contains sequences of lowercase letters joined with an underscore and false otherwise""
%\end{lstlisting}
%
% \begin{lstlisting}[style=mystyle]
% def sort(l):
%     """Sort a list of integers""
% \end{lstlisting}
Instead of directly displaying a list of plausible suggestions, our framework \tappProblem{} would query the user with a question:
\begin{lstlisting}[style=mystyle-simple]
text_lowercase_underscore("aa_bb_cc") == True? 
\end{lstlisting}

% \begin{lstlisting}[style=mystyle-simple]
% Did you mean sort([3, 2, 4]) == [2, 3, 4]? 
% \end{lstlisting}
 Let us assume that the user answers 'no', since they expect only two sequences of lowercase letters, joined by one underscore (based on the reference implementation that accompanies the problem).
The workflow would likely query the user again with the following question:
\begin{lstlisting}[style=mystyle-simple]
text_lowercase_underscore("aa_bb") == True? 
\end{lstlisting}
% \begin{lstlisting}[style=mystyle-simple]
% Did you mean sort([3, 2, 4]) == [4, 3, 2]? 
% \end{lstlisting}
If the user says 'yes', then the system would output the list of approved tests, as well as a set of semantically ranked code suggestions that are consistent with those tests. 
Once the user chooses a suggestion from such a list (ideally the top ranked one), it would generate code along with accompanying tests. 
\begin{lstlisting}[style=mystyle-simple]
def text_lowercase_underscore(text):
    return True if bool(re.search(r'^[a-z]_[a-z]+$', text)) else False

def test_text_lowercase_underscore():
    assert text_lowercase_underscore("aa_bb")== True

test_text_lowercase_underscore()
\end{lstlisting}

In the case of LLM-based code generation, the generated tests not only help make natural language intent more precise, prune incorrect suggestions generated by the LLM, but can also serves as debugging aid for remaining suggestions and regression tests for future code edits~\cite{copilot_maps_2022}. 

% \begin{lstlisting}[style=mystyle]
% def sort(l):
%     """Sort a list of integers"""
%    sorted(l, reverse=True)

% def test_sort():
%     assert sort([3, 2, 4] == [4, 3, 2]

% test_sort()
% \end{lstlisting}

%\newcommand{\testDrivenUiDAlgo}{{\it TestDrivenIntentDiscoveryWorkflow}}
\newcommand{\testDrivenUiDAlgo}{{\it InteractiveTestDrivenCodeGen}}

While the proposed framework appears intuitive, the utility of the interactive framework is contingent upon the cost-benefit trade-off of the overhead of user interaction versus the generation of accepted tests and their benefit on pruning and ranking (or perhaps repair) of code suggestions. 
This requires objective metrics to evaluate the cost and benefit of such an approach. %\shuvendu{Make sure we are not arguing against subjective metrics for user-study}.

In this paper, we contribute by studying the \tappProblem{} workflow through the following steps:
\begin{enumerate}[nosep]
\item First, we describe an abstract algorithm \testDrivenUiDAlgo{} for  \tappProblem{} that can (a) leverage off-the-shelf LLMs for generating seed tests, and (b) is parameterized by various well-specified components for pruning, mutating and ranking tests. 
Our approach is both domain-agnostic and programming-language agnostic by leveraging off-the-shelf LLMs for generating tests and rely only on {\it runtime feedback} to prune, mutate and rank tests and code. 
\item Next, we implement \testDrivenUiDAlgo{}  in a tool \TOOL{} and evaluate several heuristics across our metrics on two Python programming datasets: MBPP and HumanEval. 
%\shuvendu{Fold it into the evaluation} We describe a way to {\it simulate user response} with high-fidelity and establish a set of {\it offline metrics} that can enable us to automatically evaluate different solutions on a given benchmark  without requiring a user in the loop. 
\item Finally, we demonstrate empirically that each component of \testDrivenUiDAlgo \space (namely, interaction, prompts, mutation, ranking) contributes to improving the effectiveness of the approach over a purely LLM-based baseline approach.
Our best algorithm improves the \passk{1} code generation accuracy metric by $22.49\%$ with a {\it single} user query, and by $37.71\%$ with 5 user queries for MBPP, and by $24.79\%$ with one user query, and by $53.98\%$ with 5 user queries for HumanEval.
Second, we can generate a unit test consistent with the user intent within an average of $1.7$ user queries for $87.12\%$ of the examples in MBPP and $1.5$ user queries for $95.73\%$ of HumanEval. 
%\shuvendu{Check against new numbers}

%Our best algorithm improves the \passk{1} code generation accuracy metric from $48.24\%$ to $70.73\%\ (+22.49\%)$ with a {\it single} user query, and up to $85.95\%$ $(+37.71\%)$with up to 5 user queries for MBPP, and from $30.49\%$ to $55.28\%\ (+24.79\%)$ with one user query, and up to $84.47\%$ $(+53.98\%)$ with up to 5 user queries for HumanEval.
%Second, we can generate a unit test consistent with the user intent within an average of $1.7$ user queries for $87.12\%$ of the examples in MBPP and $1.5$ user queries for $95.73\%$ of HumanEval. \shuvendu{Check against new numbers}
\end{enumerate}

%\shuvendu{Highlight with CodeGen/Polycoder experiment that we can improve any code generator, with this emergent test generation ability of LLMs}
%For MBPP, with the best algorithm, we are able to improve pass@1 metric of correct code generation from X\% to Y\% using just a single feedback, and to Z\% using three feedbacks. On the other hand, pass@1 improves from X\% to W\% if we continue the user interaction until a test is confirmed by a user, and incurs x interactions on average. 
%Additionally, we  establish that there is significant room to improve current algorithms given the best performance an ideal algorithm can have \shuvendu{downplay}. 
%We also establish that the baseline for a purely neural approach to test generation (Codex), and illustrate significant improvements using new execution-based test mutation and test and code ranking. 
%We believe that the problem of TDUIF offers a rich area of research to not only leverage existing models and user interaction to generate correct code (i.e., code with some guarantees albeit weak), but also develop new neural models for correct code generation. 

\section{Workflow and Problem Formulation}

In this section, we first provide some background on Large Language Models and outline the workflow for leveraging test generation and user feedback to disambiguate and refine user intent. 
Next we define intuitive metrics to evaluate different approaches that implement the workflow. 
Finally, we provide an algorithm to describe the workflow.

\subsection{Background: Large Language Models for Code}
\label{sec:llm}

Autoregressive language models based on the GPT-3 architecture~\cite{gpt_3_2020} such as \emph{Codex}~\cite{codex_2021} and \emph{PolyCoder}~\cite{polycoder_2022},  predict \emph{the probability of a token} given the previous tokens, \emph{i.e.} the input string (or \emph{prompt}) of restricted length that is passed to the LLM. 
The left-to-right nature of these models makes them  useful for program generation tasks, such as code completion. 
Given a \emph{code prompt}, that may include a natural language description and/or input-output examples along with any other relevant context, a LLM generates a set of candidate completions that can be used for code completions. 

\subsection{High-level Workflow}
\label{subsec:workflow}

\newcommand{\yesResponse}{{\sc Yes}}
\newcommand{\noResponse}{{\sc No}}
\newcommand{\dnResponse}{{\sc Undefined}}

\begin{figure}[t]
    \centering
    \includegraphics[width=0.9\linewidth]{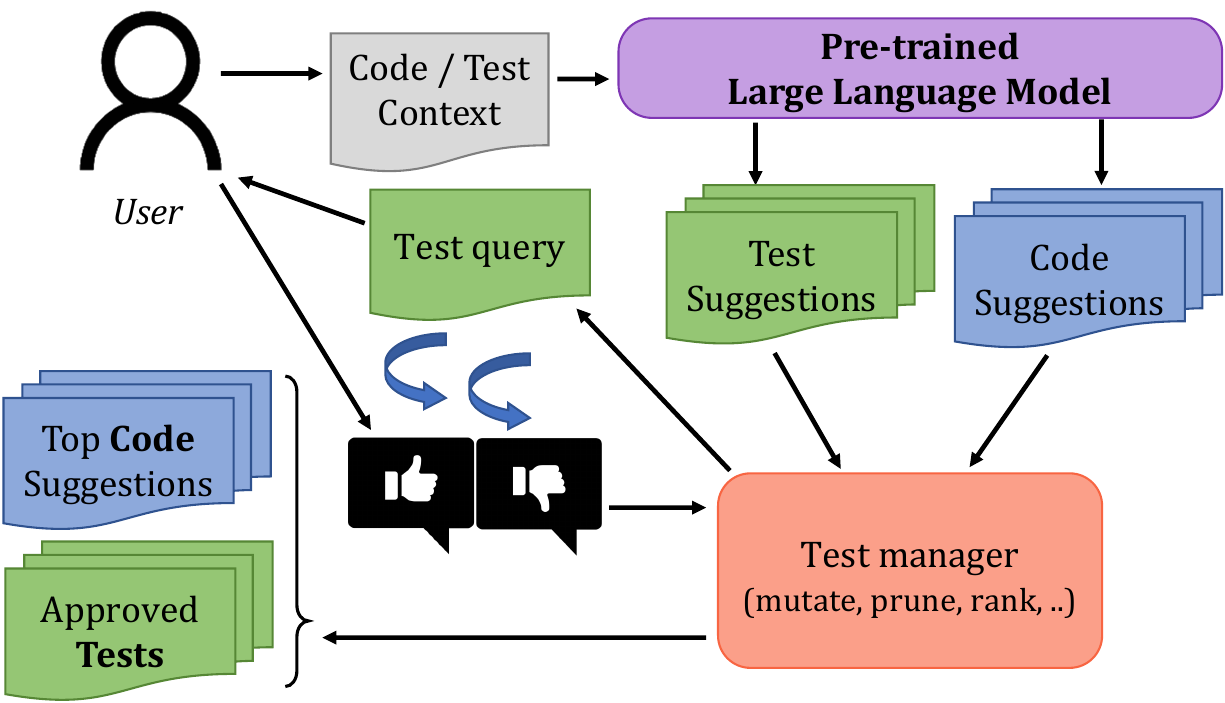}
    \caption{Workflow for interactive test-driven code generation (\tappProblem{}). }
    \label{fig:tappy-approach}
\end{figure}

Figure~\ref{fig:tappy-approach} describes the high-level workflow of {\it Interactive Test-Driven Code Generation} (\tappProblem{}).

\begin{enumerate}[nosep]
    \item The human user requests the agent for completing a function body given the prefix in a file, a natural language description and the function header/signature containing method name, parameters and returns.
    \item The agent generates a set of candidate code and test suggestions by prompting (possibly different) LLMs and possibly mutating them using {\it purely runtime} techniques.
    \item The agent chooses a test and queries the user asking if a test is consistent with the user intent.
    \item The user responds either \yesResponse{}, \noResponse, or \dnResponse{} to each of the queries from the agent.
    \item The agent leverages the user response to prune, rank and mutate the existing set of code and test suggestions. 
    \item Once the interaction terminates, the agent outputs (a) a set of tests that the user has approved, and (b) a ranked list of code suggestions that are consistent with the user responses.
\end{enumerate} 

We note a few aspects of the  workflow:
First, the proposed workflow is language and domain agnostic, as it only relies on LLMs that are increasingly capable of generating code and test suggestions in multiple popular programming languages. 
It only relies on {\it runtime execution feedback} to prune, rank and mutate suggestions that are not language-specific. 

Second, we allow a 3-valued possible user responses, including \dnResponse{}, since there are many cases when a test outcome is ill-defined in the context of the intended functionality. 
For example, when  presented with a test that violates the {\it precondition} of the desired function (e.g., \texttt{assert (SqRoot(-4) == -2)} for a square-root function that is undefined on negative numbers), it is desirable to respond \dnResponse{}. 
As another example, if the test has parse errors (e.g., \texttt{assert (foo(}), the question of the test being consistent with the user intent is not well defined. 
Finally, a {\it flaky test} that depends on non-determinism within a function may not have a unique answer (e.g., \texttt{assert (CurrentDayOfWeek() == Sunday)})~\cite{flaky-tests}.

Finally, note that our framework seeks only 1 of 3 possible user responses instead of richer feedback where, for example, the agent prompts the user to provide the desired output for a given input value~\cite{jha-icse-10}\footnote{Note that for functions returning Boolean values such as our running example in Section~\ref{sec:intro}, the distinction does not apply.}. We believe this is not only more lightweight even when a test consists of an input-output example, but the framework also generalizes well for richer tests and specifications. For example, tests of stateful APIs comprises of a test-prefix as input and the output oracle consists of a non-trivial predicate (e.g., checking non-emptiness of a stack). 
Similarly, our framework generalizes to richer symbolic postconditions where a formula tightly couples the input and the output state (e.g., saying that the output array is a permutation of the input array for a sorting function). 

\newcommand{\passat}[2]{{\tt pass@{#1}@{#2}}}
\newcommand{\passatbase}[1]{{\tt pass@{#1}}}
\newcommand{\acceptat}[1]{{\tt accept@{#1}}}
\newcommand{\positionCorrect}[1]{\tt PositionCorrect@{#1}}
\newcommand{\posCorrTests}{\tt PositiveTests}
\newcommand{\fractPosQ}[1]{\tt NumTests@{#1}}
\newcommand{\precision}[1]{\tt Precision@{#1}}
\newcommand{\numtries}{{\tt NumQueriesToAccept}}

\begin{table}[t]
\caption{Code and test generation metrics to evaluate different approaches to \tappProblem{}.}
\label{tab:metrics}

\begin{tabular}{ c|l } 
 \hline
{\bf Metric} & {{\textbf{Meaning}}} \bigstrut\\ \hline

\passat{k}{m} & Syntactic sugar for (possibly ranked) \passatbase{k} metric \bigstrut[t]\\ 
& after $m$ user queries \bigstrut[b]\\ \hline
\acceptat{m} & At least one of the proposed tests is consistent \bigstrut[t]\\
& with the user-intent after $m$ queries \bigstrut[b]\\ \hline

\end{tabular}

\end{table}

\subsection{Metrics}

Table~\ref{tab:metrics} describes two intuitive metrics to evaluate and compare the effectiveness of different approaches to \tappProblem{} over a benchmark set.
For evaluating the {\it correctness of the generated code suggestions}, we appeal to the popular metric  \passatbase{k} for evaluating the quality of code-generation by LLMs with respect to hidden tests~\cite{codex_2021}.
A code suggestion is correct if it passes all the hidden tests, and \passatbase{k} determines the expected value of choosing at least one correct code suggestion within all possible samples of size $k$. 
We define the syntactic sugar \passat{k}{m} to denote the \passatbase{k} for the code suggestions after $m$ user queries (where \passatbase{k} is the same as \pass{k}{0}).

For evaluating the {\it correctness of the generated test suggestions}, we introduce the metric $\acceptat{m}$, where $m$ denotes the number of test queries to the user. 
For a given example, the metric $\acceptat{m}$ equals 1 if at least one of the $m$ suggested test queries is consistent with the user intent, and equals 0 otherwise.

Observe that the \passat{k}{m} metric also indirectly serves to measure the quality of generated tests, by favoring tests that better prune incorrect codes.

\newcommand{\queryllm}{{\it QueryLLM}}
\newcommand{\codegen}{{\it CodePrompt}}
\newcommand{\testgen}{{\it TestPrompt}}
\newcommand{\staticMutateTests}{{\it SyntacticMutateTests}}
\newcommand{\dynMutateTests}{{\it DynMutateTests}}
\newcommand{\pruneTests}{\it PruneTests}
\newcommand{\staticPruneTests}{{\it PruneTestsStatic}}
\newcommand{\dynPruneTests}{{\it PruneTestsDyn}}
\newcommand{\satisfiesUserIntent}{{\it SatisfiesUserIntent}}
\newcommand{\posTests}{T^+}
\newcommand{\negTests}{T^-}
\newcommand{\totalTests}{T^*}
\newcommand{\nil}{\it nil}

\newcommand{\maxU}{\it MaxRounds}
\newcommand{\rankTests}{{\it RankTests}}
\newcommand{\rankCodes}{{\it RankCodes}}
\newcommand{\stopCriteria}{{\it StoppingCriteria}}
\newcommand{\compose}[2]{#1 \otimes #2}

\newcommand{\exec}[2]{{\it exec}(#1, #2)}
\newcommand{\passexec}{{\sc Pass}}
\newcommand{\failexec}{{\sc Fail}}
\newcommand{\nontermexec}{{\sc NonTerm}}

\newcommand{\programs}{{\it Prog}}
\newcommand{\tests}{{\it Test}}
\newcommand{\funcsig}[1]{#1_{\it sig}}
\newcommand{\funcbody}[1]{#1_{\it def}}
\newcommand{\prfx}{{\it prfx}}

\subsection{Abstract Algorithm}
\label{sec:algorithm}
Although the workflow is general enough to apply to any program and test framework, we make the workflow \tappProblem{} precise in the setting of a class of simple programs containing a single function to be completed.
We also restrict a test to be an {\it input-output} pair $(i,o)$.
A function $f$ satisfies a test $(i, o)$ if and only if the result of executing $f$ on $i$ terminates with a unique value $o$, i.e., $f(i) = o$.

\begin{figure}[t]
    \centering
    \includegraphics[width=\linewidth]{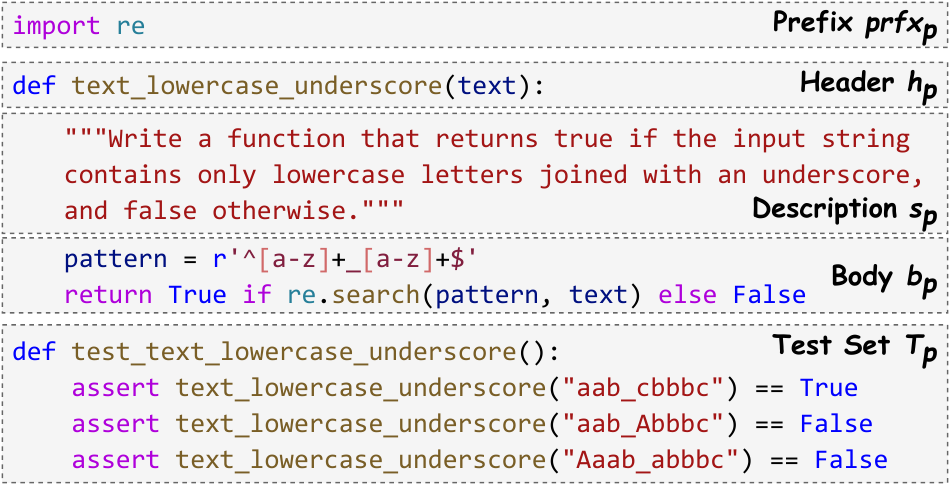}
    \caption{A Python program \emph{p} and the provided test set $T_p$. We refer to the tuple $(h_p, s_p, b_p)$ as the reference solution $f_p$.}
    \label{fig:mbpp-sample}
\end{figure}

\begin{defn}
\label{def:program}
A {\it program} $p$ is a tuple $\langle \prfx_p, s_p, h_p, b_p, T_p\rangle$, where $\prfx_p$ is a prefix that may contain definitions of other global variables and imports, $s_p$ is a natural language string description, $h_p$ is the function header or signature, $b_p$ is the body of the function and $T_p$ is a set of unit tests. 
\end{defn}
\autoref{fig:mbpp-sample} gives an example of such a program $p$ expanding our running example.

\newcommand{\llmname}{\mathcal{M}}

Algorithm~\ref{alg:abstract-tappy} describes the workflow informally sketched previously in Figure~\ref{fig:tappy-approach}.
The {\it abstract} algorithm is parameterized by a number of components that are \underline{underlined}. 
It takes as inputs the prefix $\prfx$ in a file containing imports and other global variables, the natural language description of intent $s$, the signature/header of a function $h$ including the function name and parameters.

The algorithm also takes a threshold $\maxU$ for the number of user interaction rounds. 
Once terminated, the algorithm returns a set of tests approved by the user $\posTests$ as well as a ranked list $G$ of candidate implementations of $f$ that satisfies all the tests in $\posTests$.

The algorithm starts off by generating sets of code and test suggestions in to the variables $G$ (line~\ref{line:codeprompt}) and $U$ (line~\ref{line:testprompt}) respectively.
The quality of these sets will depend on the choice of the large language model $\llmname{}$ as well as the code and test {\it prompts} constructed from the problem description\footnote{Although there exists powerful automated test generation tools such as Randoop~\cite{randoop-icse07} for Java and C\#, our algorithm uses LLMs  to generate the initial seed tests as (a) we do not have the body of the method under test, and (b) these techniques are language-specific.}. 

We allow the test generation prompt to take the set of generated codes in $G$, to possibly improve the prompt. 
The algorithm maintains the invariant that the final set of code suggestions returned to the user is always a (possibly empty) subset of the initial set of suggestions in $G$, and never prunes a correct  (as defined by the hidden tests)  code suggestion. 
On the other hand, we allow the set of tests in $U$ to be modified or augmented through both syntactic (line~\ref{line:staticmutate}) and dynamic mutation (line~\ref{line:dynmutate}) techniques using $\staticMutateTests$ and $\dynMutateTests$ components respectively.
Notice the dynamic mutation technique $\dynMutateTests$ takes the set of code suggestions in $G$ as an input (in addition to $U$).

Finally, iterating in a loop (lines~\ref{line:while} to~\ref{line:endwhile}) until the stopping criteria is satisfied or the set of candidate tests in $U$ is empty, ranking the tests in $U$ using the method $\rankTests$ (line~\ref{line:testrank}) and queries the user with the top-ranked test (line~\ref{line:queryuser}).
If the user accepts the test, the set $\posTests$ is updated (line~\ref{line:postest1}), and any code suggestion in $G$ that disagrees with the test is pruned away (line~\ref{line:posprune}).
Conversely, if the user rejects the test, then any code suggestion in $G$ that agrees with the test is pruned away (line~\ref{line:negprune}).
No action is taken if the user responds with \dnResponse{}.
Finally, the code suggestions in $G$ are re-ranked with the remaining test suggestions in $U$ (line~\ref{line:rankcode}) after the pruning.

\begin{algorithm}[t]
\caption{\testDrivenUiDAlgo{}}
\label{alg:abstract-tappy}{\footnotesize
\begin{algorithmic}[1]
\Require Prefix $\prfx$, description $s$, header $h$ of a function $f$

\Require Maximum number of interactions $\maxU$
\Ensure A ranked (possibly empty) list of candidate implementations for $f$ $G$, 
\Ensure A (possibly empty) set of user-approved tests $\posTests$ 
\Ensure $f'(i) == o$ for each $f' \in G$ and $(i, o)$ in $\posTests$

\State $G \leftarrow \queryllm(\underline{\llmname{}}, \underline{\codegen}(\prfx, s, h))$ \Comment{Query LLM for codes} \label{line:codeprompt}
\State $U \leftarrow \queryllm(\underline{\llmname{}}, \underline{\testgen}(\prfx, s, h, G))$ \Comment{Query LLM for tests} \label{line:testprompt}
\State $U \leftarrow \underline{\staticMutateTests}(U)$ \Comment{Mutate tests statically} \label{line:staticmutate}
\State $U \leftarrow \underline{\dynMutateTests}(U, G)$ \Comment{Mutate tests using dynamic execution} \label{line:dynmutate}

\State $\posTests, k \leftarrow \{\}, 0$

\While{$k \leq \maxU$  and $|U| > 0$} \label{line:while}
   \State $U \leftarrow \underline{\rankTests}(U, G)$ \Comment{Rank the test suggestions} \label{line:testrank}
   \State $(i, o) \gets U.pop()$ \Comment{Remove the top ranked test} \label{line:testpop}
   \State $k \gets k + 1$ \Comment{Number of user queries}
   \State $r \gets \satisfiesUserIntent((i, o), f)$ \Comment{Query user for intent} \label{line:queryuser}
   \If{r == \yesResponse}
        \State $\posTests \gets \posTests \cup \{(i, o)\}$ \label{line:postest1}
        \State $G \gets G \setminus \{c \ | \ c(i) \neq o\}$ \Comment{Prune codes that fail the accepted test} \label{line:posprune}
    \ElsIf{r == \noResponse}

        \State $G \gets G \setminus \{c \ | \ c(i) == o\}$ \Comment{Prune codes that pass the rejected test} \label{line:negprune}
   \EndIf

   \State $G \gets \underline{\rankCodes}(G, U)$ \Comment{Rank the code suggestions} \label{line:rankcode}

\EndWhile \label{line:endwhile}
\State \Return {$G, \posTests$} 
\end{algorithmic}}
\end{algorithm}

\section{\TOOL{}}
\label{sec:ticoder}

In this section, we describe \TOOL{} (\underline{T}est-driven \underline{I}nteractive \underline{Coder}), a tool that implements the various components that are underlined in Algorithm~\ref{alg:abstract-tappy}.
For each component (such as \codegen{}, \rankTests{}), we provide several possible alternate implementations to define the space of solutions.

\subsection{Code and Test Generation Prompts}

It is well-known that the choice of prompts that determine the actual string that is fed to an LLM has a substantial impact on the quality of output~\cite{prompt_engineering_2021}.
In this section, we outline several choices for implementing the prompt generation routines \underline{\codegen{}} and \underline{\testgen{}}  for generating code and test suggestions from the problem description consisting of $(\prfx_p, s_p, h_p)$.

\newcommand{\ttt}[1]{{\tt #1}}

\begin{figure}[t]
    \centering
    \includegraphics[width=\linewidth]{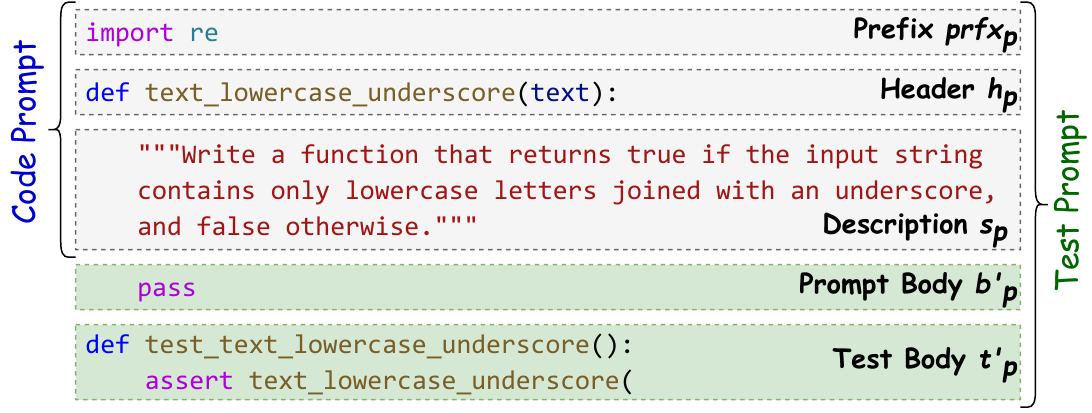}
    \caption{Example \emph{code} and \emph{test prompts} for the running example in \autoref{fig:mbpp-sample} that instantiate the algorithm $A(\prfx_p, s_p, h_p)$.}
    \label{fig:mbpp-sample-prompts}
\end{figure}

\begin{figure*}[htbp]
    \centering
    \includegraphics[width=0.97\linewidth]{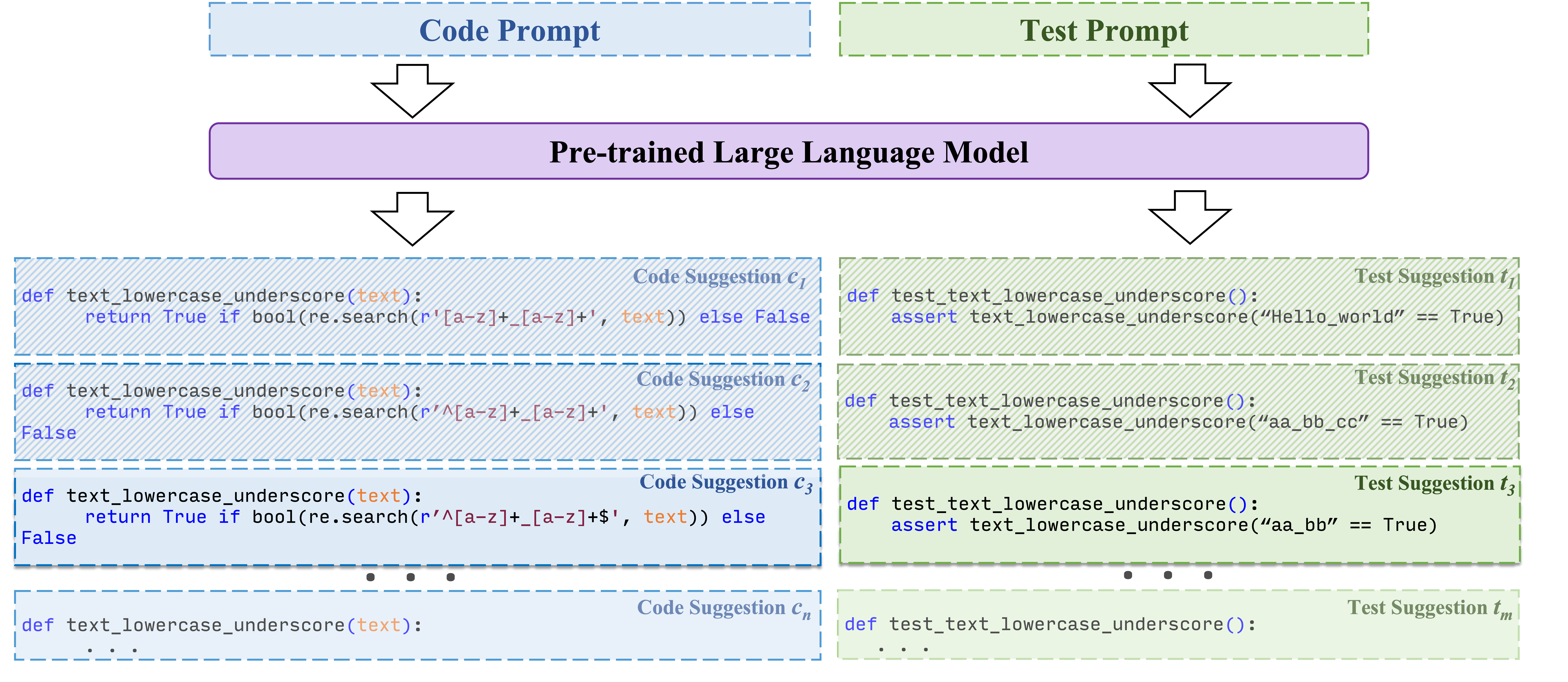}
    \caption{\emph{Code} and \emph{test suggestions} for the running example in \autoref{fig:mbpp-sample} generated from a LLM. Code suggestion $c_3$ and test suggestion $t_3$ are both \emph{correct}, while code suggestions $c_1$, $c_2$ and test suggestions $t_1$, $t_2$ are \emph{incorrect} (appear shaded), \emph{i.e.} they don't satisfy the \emph{problem prompts} in \autoref{fig:mbpp-sample-prompts}.}
    \label{fig:mbpp-sample-suggestions}
\end{figure*}

\autoref{fig:mbpp-sample-prompts} presents a possible \emph{code prompt} (in the gray boxes) that is generated by $\codegen{}(\prfx_p, s_p, h_p)$ and can be passed to a LLM to produce \emph{code suggestions} for the given problem in \autoref{fig:mbpp-sample}. 
Querying a LLM (e.g. Codex) with the code generation prompt in \autoref{fig:mbpp-sample-prompts} will result in a set of \emph{code suggestions} as shown in \autoref{fig:mbpp-sample-suggestions}. 
Code suggestion $c_3$ is a valid solution to the problem, while $c_1$ is an incorrect code suggestion (since it allows the first substring to start with an uppercase letter) and $c_2$ is also incorrect (since it allows more than one sequence of lowercase letters joined with an underscore).

There are several interesting choices for generating the test  prompts for \testgen{}. 
Given that we wish to generate a test for a function without an implementation, the problem of  \testgen{} in our setting boils down to completing the method body of $f$. 
The green boxes in \autoref{fig:mbpp-sample-prompts} show the ``Prompt Body" and the subsequent ``Test Body" that, together with the code prompt, constitute the test prompt.
We use the statement \texttt{pass} as the method body, corresponding to a placeholder implementation in Python.
The generated test suggestions (\autoref{fig:mbpp-sample-suggestions}) present the user with a set of tests.
Some of these are \emph{consistent} with the user intent ($t_3$); while  others are  \emph{inconsistent} with the user intent ($t_1$ and $t_2$). 

However, there are several other interesting possibilities for designing \testgen{}.
For instance, we can sample a code completion $b'_p$ from the set of generated code suggestions $G$ and use it as definition of $f$.
We therefore explore two options for $b_p$ in this work:
\begin{itemize}
    \item \texttt{pass}: We can instantiate $b_p$ to simply be \texttt{pass} to keep the prompt for the test generation syntactically correct, as illustrated in \autoref{fig:mbpp-sample-prompts}.
    \item {\it choose}($G$): Alternatively, we can sample a definition from a set of code suggestions $G$ and use it to instantiate the body of $f$.

    In our experiment, we choose the code suggestion that appears first in the set of code suggestions in $G$ returned by the LLM. 
\end{itemize}

\subsection{Static mutation of tests with \staticMutateTests{}}
Given an initial set of candidate tests in $U$, one can perform various syntactic mutations of a given test $t$ to yield new test cases.
For each test $t \in U$, we consider two options for statically mutating it:
\begin{itemize}
    \item In case a test has a parsing error, we consider the longest prefix of $t$ that parses.
    \item We consider the prefix of $t$ up to the first assertion; since each assertion is a point of failure, considering only one assertion maximizes the chance of a test passing. We refer to this technique as \texttt{single-assert}.
\end{itemize}

However, note that some of these decisions (such as \texttt{single-assert}) may also adversely impact the performance as it weakens the checks in the tests.

\subsection{Dynamic mutation of  tests with \dynMutateTests{}}
In addition to statically mutating tests, one can also exploit the ability to execute the tests, using the execution feedback to obtain new tests. 

Given a test $(i,o)$ and a candidate implementation $f' \in G$, we can generate an alternate test $(i, f'(i))$ by modifying the output value observed by executing $f'$ (if any). 
The intuition behind this is that if $f'$ happens to be a correct solution, then we generate at least one test that is consistent with $f'$.
We have implemented \texttt{assert-rewrite-all}, where we {\it augment} $U$ with all the tests obtained by rewriting each $(i, o) \in U$ with $(i, f'(i))$ for each $f' \in G$.

Consider a simple example, where we rewrite the test $t_2$ to $t'_2$ in \autoref{fig:mbpp-sample-suggestions},  where we mutate the test $t_2$ (\texttt{"aa\_bb\_cc"}, \texttt{True})  to  (\texttt{"aa\_bb\_cc"}, \texttt{False}) by executing the input of $t_2$ through suggestion $c_3$ to obtain a new test $t_2'$. 
Now $t'_2$ is  consistent (i.e. passes) with the correct suggestion $c_3$.
Although in this particular case (given the Boolean valued function) both $t_2$ and $t_2'$ have the same pruning ability on the current code suggestions that have a Boolean return value, accepting $t_2'$ will prune any (hypothetical) code suggestion $c_n$ that incorrectly returns an integer as output.
The mutation is more effective when the output value comes from a large space of values (e.g., strings, lists or integers).

\subsection{Ranking test suggestions using \rankTests{}}
\label{sec:technique:testranking}

After obtaining the set of tests $U$, the user is presented with a test $t \in U$, which the user answers, thus pruning away code suggestions that are inconsistent with the user's suggestions. 

Therefore, to minimize the overhead of number of user interactions, it is necessary to present tests to the user that would result in the most number of incorrect code suggestions being pruned away~\cite{jha-icse-10,le2017interactive}.
We describe various test ranking strategies starting with the performance of an optimal ranker. 

\subsubsection{Ideal} \texttt{(ideal)} 
Given the set of tests $U$, we first define the {\it optimal} (yet unrealizable) ranking policy.
This option simulates the effect of choosing each test $t \in U$ to present to the user, and the number of incorrect code suggestions that will be pruned for such a choice by executing lines~\ref{line:queryuser} to ~\ref{line:negprune} of Algorithm~\ref{alg:abstract-tappy}.
Since the Algorithm~\ref{alg:abstract-tappy} only prunes incorrect code suggestions, the above simulation would identify the test in $U$ that would maximize the pruning of incorrect suggestions, and establish an upper bound.   
However, this strategy is not realizable, as it requires the labeling of code suggestions as correct or incorrect.

We now discuss several alternative realizable approaches to implementing \underline{\rankTests{}}.

\subsubsection{Random} \texttt{(random)}
In this policy, we present the user with a test $t$ randomly chosen from $U$.
This serves as a baseline where the user is simply presented with a test suggestion without any estimate of how useful presenting that test to the user would actually be.

\subsubsection{Discriminative} \texttt{(discriminative)}
In this policy, we rank the tests from $U$ based on how well they discriminate the set of code suggestions in $G$.
If a test $t$ can discriminate between code suggestions well (i.e., splits the set of code suggestions into roughly equal halves), then it would prune away a substantial fraction of the code suggestions irrespective of the user response. 
Under the assumption that each code suggestion is equally likely to be correct or incorrect, this heuristic is likely going to yield a good test ranking strategy.  

More precisely, for each  test $t \in U$, we split the set of code suggestions $G$ into the sets $G^+_t$ and $G^-_t$ of code suggestions that pass and fail the assertions in $t$, respectively.
We then prioritize tests where the ratio of the sizes of these two set is closest to 1. 
In other words, we  rank the tests in decreasing order using the following scoring metric $s_{discr}$:

\begin{equation*}
    \begin{split}
        \textit{min} = & \texttt{min}(|G^+_t|, |G^-_t|) \\
        \textit{max} = & \texttt{max}(|G^+_t|, |G^-_t|) \\
        s_{\textit{discr}}(t) = & \left\{
            \begin{array}{lr}
                0 & \text{if } \textit{max} \text{ is }0 \\
                \textit{min} / \textit{max} & \text{otherwise}
            \end{array}
        \right\}
    \end{split}
\end{equation*}

Note that we do not consider runtime exceptions and precondition failures as part of $G^-_t$. 
Our reasoning is similar to how we define the $\satisfiesUserIntent$ predicate, where we expect the user would likely respond with \dnResponse{} to such tests. 

Consider the example in Figure~\ref{fig:mbpp-sample-suggestions}. 
Consider the two tests $t_1$ and $t_2$:
Two code suggestions \{$c_2$, $c_3$\} \ic{FAIL} on test suggestion $t_1$ while one suggestion \{$c_1$\}  \ic{PASS}, making $s_{\textit{discr}}(t_1) = \texttt{min}(1,2)/\texttt{max}(1,2) = 1/2$. Similarly, two code suggestions \{$c_1$, $c_2$\} \ic{PASS} on test suggestion $t_2$ while one suggestion \{$c_3$\}  \ic{FAIL} and $s_{\textit{discr}}(t_2) = 1/2$. Both tests $t_1$ and $t_2$ have equal \texttt{discriminative} ranking, although \texttt{ideal} ranking would choose to show $t_2$ as it prunes more incorrect code.
All code suggestions in this example \ic{PASS} on test $t_3$ making $s_{\textit{discr}}(t) = 0$. In this example, it is clear that showing $t_3$ would prune away the least number of incorrect code suggestions, and would not be chosen.

\subsection{Ranking code suggestions using \rankCodes{}}
\label{sec:rankcode}
Finally, our goal is to present the user with a ranked list of code suggestions in $G$.
We currently define a single code ranking strategy (\texttt{passing-tests}) that uses the tests in $U$ to determine an ordering on $G$ as follows:
\begin{itemize}
    \item Each generated code $c \in G$ is executed with each test $t \in U$ and gets assigned as a score \emph{the number of passing tests} $d_c$. The codes are then ranked based on the decreasing order of $d_c$. 
\end{itemize}

Other variations of clustering and ranking code suggestions using tests have also been previously explored in recent works~\cite{codet_2022,alphacode_2022}, but currently not implemented in \TOOL{}.

Following from the example in the previous section, represented in ~\autoref{fig:mbpp-sample-suggestions}, code suggestion $c_1$ passes on all tests \{$t_1$, $t_2$, $t_3$\}, code suggestion $c_2$ passes on \{$t_2$, $t_3$\} and code suggestion $c_3$ passes on \{$t_3$\}.
Our ranking would therefore rank $c_1$ highest initially in the absence of any feedback from the user.

\newcommand{\defaultConfig}{\tt Default}
\newcommand{\idealConfig}{\tt Ideal}
\newcommand{\baselinePassConfig}{\tt Baseline-PassPrompt}
\newcommand{\baselineCodeConfig}{\tt Baseline-CodePrompt}
\newcommand{\staticPruneConfig}{\tt StaticPruning}
\newcommand{\dynPruneConfig}{\tt DynPruning}
\newcommand{\regrConfig}{\tt RegressionTests}
\newcommand{\orderByDiscriminationConfig}{\tt OrderByDiscrimination}
\newcommand{\orderByAndFeed}{\tt OrderByDiscrimination-Feed}

\section{Experimental Evaluation}
\label{sec:eval}
In this section, we start with our main research questions  to evaluate different approaches and techniques for the test-driven interactive code generation problem.
We next describe the experimental setup including datasets and how we automate the experimental setup.
In Section~\ref{sec:results}, we describe our results. 

\subsection{Research Questions}
\label{sec:rq}

\newcommand{\idealTests}{{\it IdealTests}}
\newcommand{\baseline}{{\it Baseline}}
\newcommand{\idealRanking}{{\it IdealRanking}}

\begin{enumerate}
\item {\bf RQ1}: How does the interactive workflow improve the accuracy of code suggestions?
\item {\bf RQ2}: How does the correctness of generated code and tests improve with the number of user queries?
\item {\bf RQ3}: How do each of the design decisions affect the metrics (ablation study)? 
\end{enumerate}

\subsection{Dataset}
\label{sec:datasets}

We use two Python programming datasets for our evaluation, including the {\it sanitized} version of the \emph{MBPP dataset} ~\cite{google_llm_2021}, dataset from Google, and the \emph{HumanEval dataset}, introduced in the Codex paper~\cite{codex_2021}, to answer the research questions.
\emph{MBPP} consists of 427 $\langle \prfx_p, s_p, h_p, b_p, T_p \rangle$ tuples and \emph{HumanEval} of 164 tuples as per Definition \ref{def:program}, where $b_p$ is the ground truth definition of the corresponding function.
One example of such a tuple has been discussed in \autoref{fig:mbpp-sample}. We modify the original \emph{HumanEval} dataset to remove any (non-hidden) input-output examples that are included in the docstring; in the presence of such examples in the docstring, one can write simple rules to propose a test that is guaranteed to satisfy the user intent as well as prune many incorrect code suggestions.

\subsection{Experimental setup and tools}
\label{setup}

For all experiments, we use Open AI's Codex \texttt{code-davinci-002} model for inference only exposed through APIs. \footnote{Access to this model was removed to the public by OpenAI in March 2023, but continues to be made free and available to researchers upon request.}
In each case, we query the Codex model for \emph{100 code suggestions,} with a temperature of 0.8 and a top $p$ of 0.95.
Intuitively, a temperature closer to 1 allows LLMs (including Codex) to provide a more diverse set of solutions, whereas a temperature closer to 0 forces LLMs to only generate fewer solutions with the highest confidence. 
The maximum code generation length is 300 tokens. 
Additionally, we query the Codex model for \emph{50 test suggestions} using the same parameters as before.
Results are reported using the \emph{expected values} of either \passat{k}{m} or \acceptat{m} metrics described in Section~\ref{subsec:workflow} over all examples in each dataset.
Further, in order to account for the non-determinism of Codex, we only query Codex once to generate the initial  code and test suggestions into a \emph{cache} of Codex responses and refer to the same cache for all experiments.

We have implemented our approach in \besttool{} and we compare the performance of our approach with the Codex's standard code generation, without including any user interaction.
We also consider a \baseline{} version of \besttool{}, where the tests to be presented to the user are generated by Codex with ${\it TestGenPrompt}=\texttt{pass}$ with no further mutation, pruning or ranking while also disabling code ranking.
Finally, we also compare against our implementation of a recent related work \emph{CodeT}~\cite{codet_2022}, which also exploits LLM-generated tests to improve the quality of code generation without any user interaction. 
In a nutshell, \emph{CodeT} ranks a code (respectively, a test) by the number of tests (respectively, codes) it satisfies. 

We also consider a few settings that help us establish an upper bound on the performance of any solution:
\begin{enumerate}
\item First, we define \idealTests{} where the tests presented to the user only consist of the hidden tests from $T_p$, ranked using the \idealRanking{}. This configuration defines an upper bound for our metrics.
\item Second, we consider \idealRanking{}, where the set of tests generated in $U$ are ranked in the ideal manner described in Section~\ref{sec:technique:testranking}. 

Although unrealizable, it helps determine the optimal ranking within a set of generated tests. 
\end{enumerate}

For the default \besttool{}, we set:

${\it TestGenPrompt}$=\texttt{pass}

${\it StaticMutateTests}$=\texttt{single-assert} 

${\it DynMutateTests}$ =\texttt{assert-rewrite-all}

${\it TestRanking}$ = \texttt{discriminative} 

${\it CodeRanking}$ =\texttt{passing-tests}

We chose this configuration as default empirically as it performs the best on the \passat{1}{1} metric on the MBPP dataset.

\begin{figure}[t]
    \centering
    \begin{subfigure}[t]{0.44\textwidth}
        \centering
        \includegraphics[trim={0cm 0.38cm 0cm 0cm}, clip, width=\linewidth]{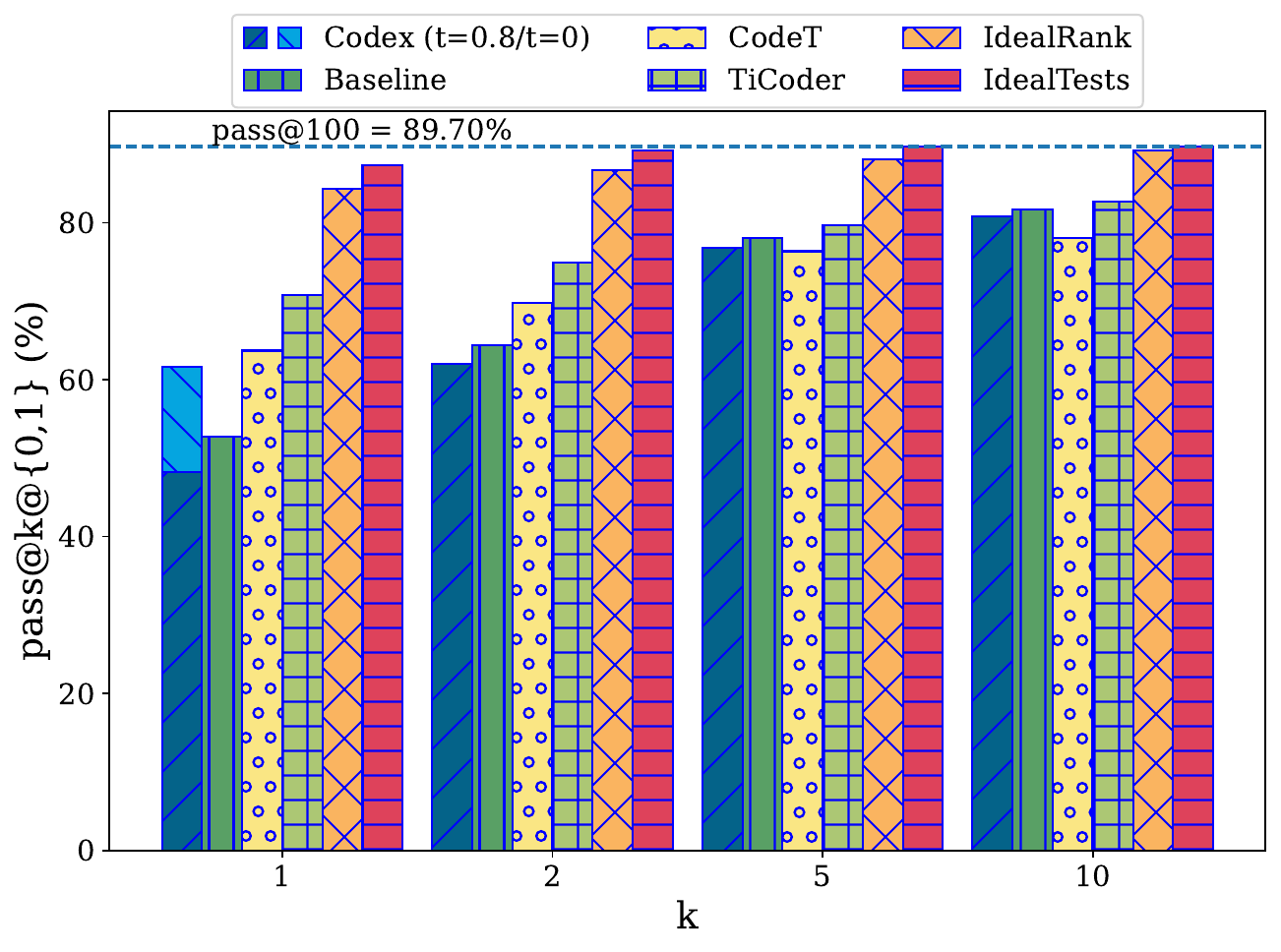}
        \caption{\emph{MBPP} dataset} 
        \label{fig:rq1_mbpp}
    \end{subfigure}
    
    \begin{subfigure}[t]{0.44\textwidth}
        \centering
        \includegraphics[trim={0cm 0.38cm 0cm 0cm}, clip, width=\linewidth]{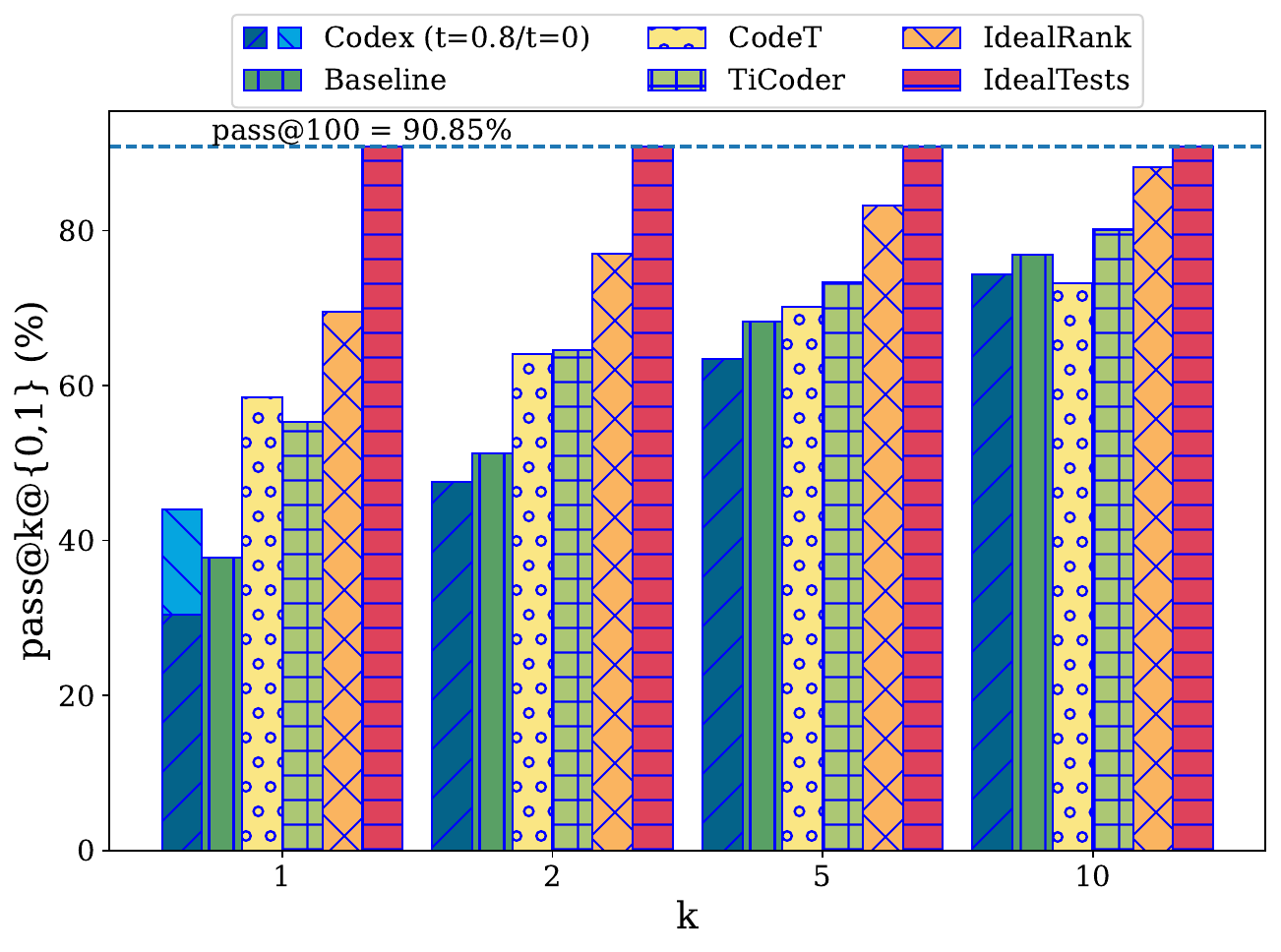}
        \caption{\emph{HumanEval} dataset}
        \label{fig:rq1_human_eval}
    \end{subfigure}
    \caption{Code accuracy results.}
\end{figure}

\subsection{Automated evaluation}
\label{sec:simulating-users}

In practice, our proposed workflow requires real-time user response to determine if a generated test is consistent with the user's intent (i.e. \satisfiesUserIntent{} in Algorithm~\ref{alg:abstract-tappy}). Therefore, evaluating \TOOL{} {\it offline} with large-scale benchmark datasets is largely impractical. 
This is a common challenge faced by various tools and interaction models that require user input. To address this issue, and inspired by the framework of {\it Oracle-Guided Inductive Synthesis} (OGIS)~\cite{jha-icse-10,le2017interactive,jha-acta-17}, we propose the use of the oracle, i.e. the reference code implementation in each benchmark dataset, as a proxy for the user response. 
We use the reference implementation $b_p$ as an {\it oracle} to answer if a test $(i,o)$ is consistent with the user intent. 
In other words, we assume that the intent of the user is precisely the semantics of the (hidden) reference implementation $f_p$ for all {\it  terminating executions} of $f_p$.

We define the use of the oracle as a proxy for the user response as follows:

\begin{defn}\label{def:usersim}
For a deterministic function $f$ in a program $p$ with a reference implementation $f_p$ (comprising of $h_p$ as header and $b_p$ as the body) of $f$, and a test (or an input/output example) $(i, o)$, $\satisfiesUserIntent((i, o), f)$ returns
\begin{enumerate}[nosep]
    \item  \yesResponse{}, if $f_p(i)$ terminates and produces the output $o$. 
    \item  \noResponse{}, if $f_p(i)$ terminates and produces an output $o' \neq o$.
    \item \dnResponse{}, denoting syntax errors, runtime exceptions, or infinite loops.
   
\end{enumerate}
\end{defn}

\section{Results}
\label{sec:results}

\subsection{RQ1: Code suggestion accuracy}

To answer RQ1, we compare the accuracies of \baseline{}, \idealTests{}, \idealRanking{}, and \besttool{} (default option) restricted to the case of a single user query. 
We use the \pass{k}{m} metric, where $k \in \{1, 2, 5, 10\}$ and $m = 1$.
We also compare the tools with Codex without any user interaction ($m = 0$).
Additionally, to get the best possible result from Codex, we also show Codex$_{t=0}$, where we query Codex  for 1 suggestion with temperature 0.
We do not show results for Codex$_{t=0}$ for $k \ge 1$ since we query for only one suggestion. 
Finally, recall that CodeT doesn't include user interaction~\cite{codet_2022} and therefore we only report \passat{k}{0} numbers here, after applying their code and test ranking strategies.

\autoref{fig:rq1_mbpp} presents the results for \emph{MBPP} and \autoref{fig:rq1_human_eval} the results for \emph{HumanEval}. We observe that \TOOL{} has a \pass{1}{1} of 70.73\% for \emph{MBPP}, improving over the baseline Codex (48.24\%) by non-trivial percentage.
It also outperforms other baselines such as 
Codex$_{t=0}$ (61.59\%), \baseline{} (52.69\%) and CodeT (63.70\%) .
Similarly, \autoref{fig:rq1_human_eval} also shows that \TOOL{} has a \pass{1}{1} of 55.28\% for \emph{HumanEval}, again improving over Codex baseline (30.49\%).
It also outperforms Codex$_{t=0}$ (44.02\%) and \baseline{} (37.80\%), but falls a bit behind CodeT (58.54\%).
In Table~\ref{tab:rq3_human_eval}, we do note that an alternate non-default configuration of \TOOL{} with $\passat{1}{1}$ of 59.62\% outperforms CodeT as well. 
\TOOL{} continues outperforming \baseline{} and Codex for $k \in \{2, 5, 10\}$ for both datasets and additionally outperforms CodeT for all these metrics.

However, as expected, it falls short of the optimal \idealTests{}, with the \pass{1}{1} metric trailing by 16.62\% and 35.57\% behind for \emph{MBPP} and \emph{HumanEval} respectively. 

The comparison with \idealRanking{} illustrates that there is  room to improve the test ranking option itself, showing up to 13.58\% and 14.29\% potential improvement for \passat{1}{1} respectively for the two benchmarks.

Finally, note that the performance of \idealTests{} and \idealRanking{} are comparable for MBPP as well as HumanEval for larger values of $m$, but \idealRanking{} is always slightly lower (as expected).
This shows that our test generation strategies that include tests generated by Codex along with the static/dynamic mutations can for the most part generate high-quality disambiguating tests.

\subsection{RQ2: Impact of user interaction }

\begin{figure}[t]
    \centering
    \begin{subfigure}[t]{0.88\linewidth}
        \centering
        \includegraphics[trim={0.45cm 0.45cm 0.55cm 0.30cm}, clip, width=\linewidth]{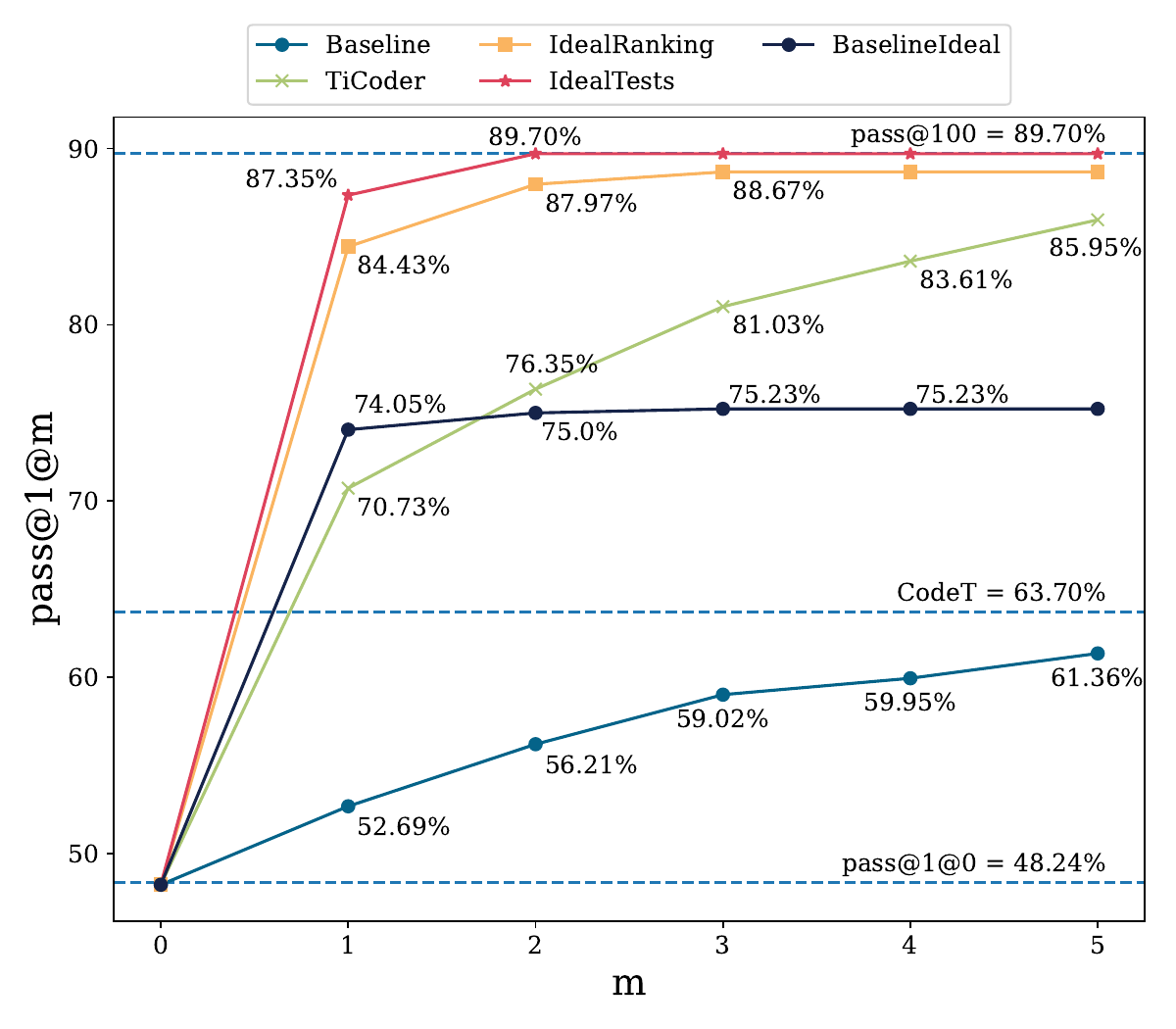}
        \caption{\emph{MBPP} dataset.}
        \label{fig:rq2_mbpp}
    \end{subfigure}
    \begin{subfigure}[t]{0.88\linewidth}
        \centering
        \includegraphics[trim={0.45cm 0.45cm 0.55cm 0.30cm}, clip, width=\linewidth]{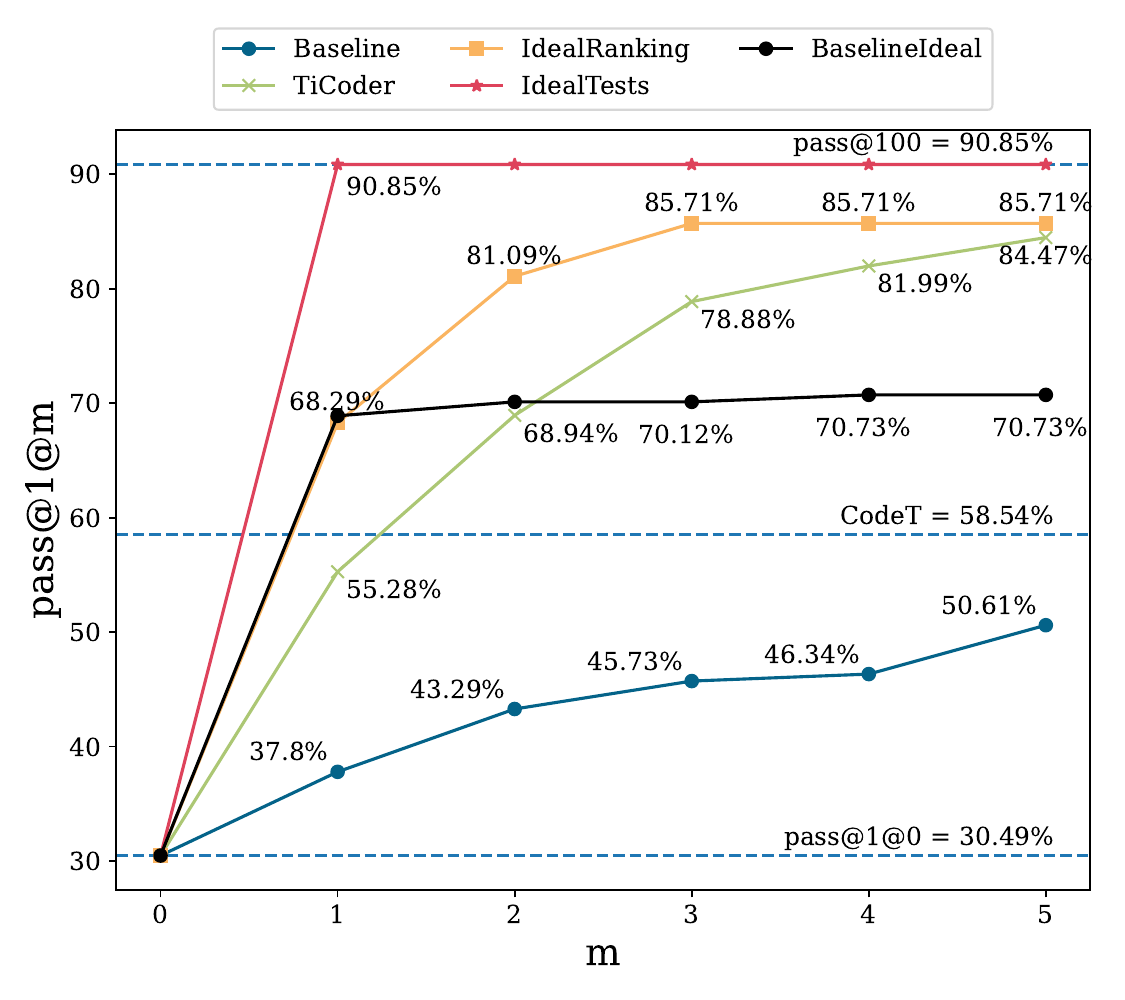}
        \caption{\emph{HumanEval} dataset.}
        \label{fig:rq2_human_eval}
    \end{subfigure}
    \caption{User interaction results.}
\end{figure}

To answer RQ2, we evaluate the four configurations along with CodeT as a baseline.
In addition, we also evaluate a new configuration {\it BaselineIdeal} that we discuss later.

We show the results for \pass{1}{m} in \autoref{fig:rq2_mbpp} for MBPP and \autoref{fig:rq2_human_eval} for HumanEval. 
In all cases, increasing the limit of the maximum number of queries increases the performance.
However, this increase is very slight for the \baseline{}, while it is considerable for \TOOL{}. 
Note that \idealTests{} achieves the highest possible performance matching the \texttt{pass@100} value after using all the hidden tests (3 tests for \emph{MBPP} and 1 test \emph{HumanEval}), while \idealRanking{} closely follows it.
With 5 interactions, the accuracy of \TOOL{} approaches the accuracy of the \idealRanking{}, signifying that there is a good benefit to offset the interaction cost. 
Improvements to the ranking policy can result in further improvements  for \besttool{} and can require fewer user interactions.

To understand if our test mutation techniques improve the pool of tests over the pool of purely LLM-generated tests, we also apply optimal ranking  to only the tests generated by the \baseline{} approach (reported as  {\it BaselineIdeal}). Both Figures~\ref{fig:rq2_mbpp} and~\ref{fig:rq2_human_eval} demonstrate that our test mutation improves the accuracy by $10.72\%$ and $13.74\%$ respectively for \emph{MBPP} and \emph{HumanEval} after 5 interactions.

We additionally observe that CodeT consistently outperforms the simple \baseline{}, even though it includes no user interactions. However, \TOOL{} outperforms CodeT with only 1 user interaction in MBPP and with 2 user interactions in HumanEval.

\begin{figure}[t]
    \centering
    \includegraphics[trim={0.52cm 0.58cm 0.52cm 0.52cm}, clip, width=0.98\linewidth]{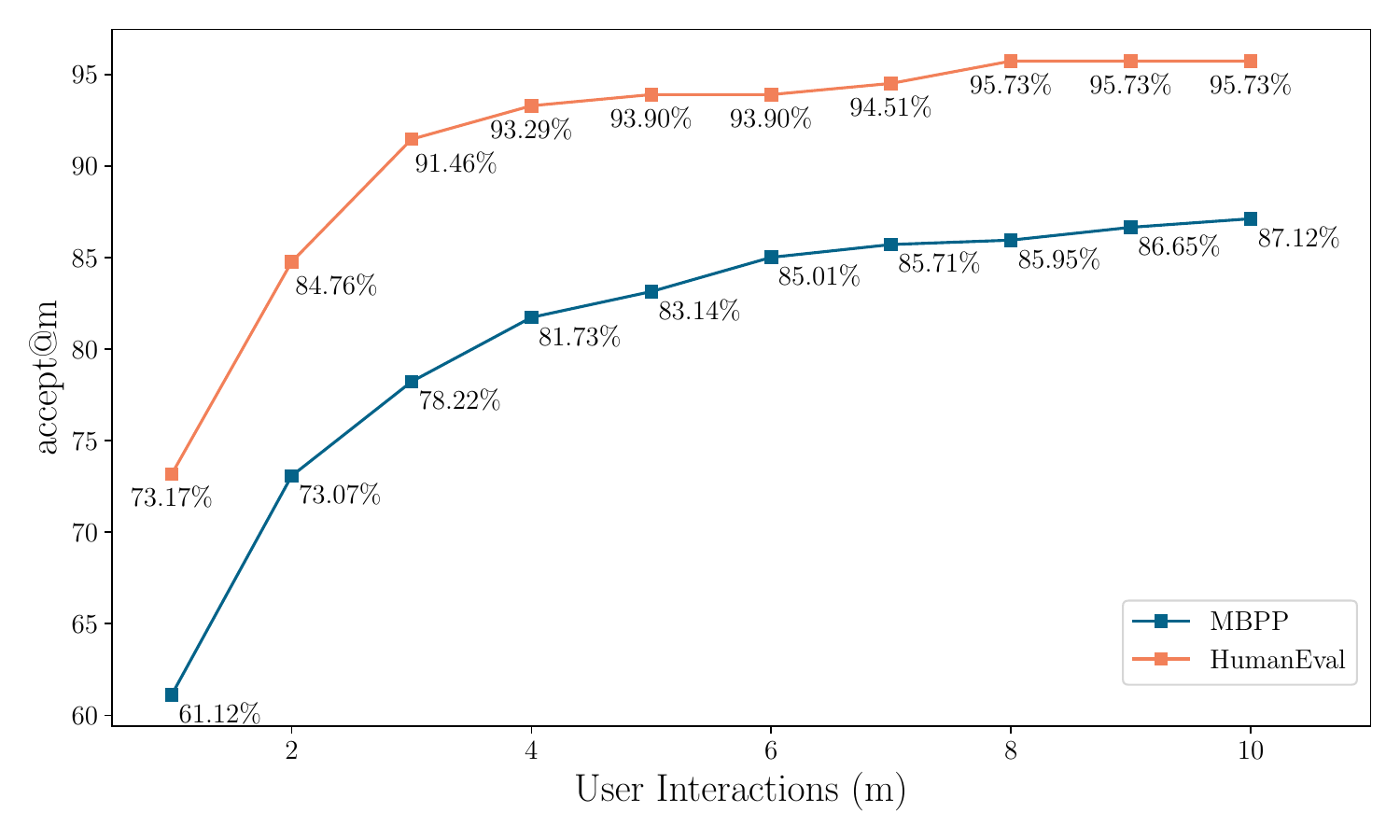}
    \caption{Plot showing fraction of examples (on y-axis) with \acceptat{m} $= 1$ for different number of user queries $m$ (on x-axis). The number of possible user queries was capped at 10.}
    \label{fig:intplot}
\end{figure}

\autoref{fig:intplot} shows the cumulative fraction of examples that produced a user-accepted test within $m$ user queries, \emph{i.e.} the expected value of \acceptat{m} for a given $m$. Additionally, it takes an average of 1.7 and 1.5 interactions with \besttool{} to find a test satisfying user intent for \emph{MBPP} and \emph{HumanEval} respectively, with 1 being the minimum, and 10 being the maximum number of interactions (we cap the maximum number of user interactions to 10).

We observe that \TOOL{} is able to propose a  test that is consistent with the user intent for 87.12\% of the examples within 10 queries for \emph{MBPP} and 95.73\% for \emph{HumanEval}, whereas the first query provides such a consistent test for 61.12\% and 73.17\% of examples respectively.  

The results demonstrate that \TOOL{} is able to propose a consistent test (that can serve as a unit test accompanying the code) in a large fraction of cases within a small number of trials. 
Moreover, these test cases are non-vacuous (i.e., not {\tt assert true}) and have good discriminative power as they prune the incorrect code suggestions and improves the \passat{1}{m} for $m \geq 1$.

\subsection{RQ3: Ablation study}

\begin{table}[t]
    \centering
    \caption{Results of the MBPP ablation studies. Metric \textit{p@k@m} stands for \textit{pass@k@m}.}
    \label{tab:rq3_mbpp}
    \resizebox{\linewidth}{!}%
    {
    \begin{tabular}{l||c|c|c||c|c}
    \hline
        {\textbf{Tool}} & \emph{\textbf{p@1@1}} & \emph{\textbf{p@2@1}} & \emph{\textbf{p@5@1}} & \emph{\textbf{p@1@2}} & \emph{\textbf{p@1@5}} \bigstrut\\ % & \emph{pass@1@*} \\
        \hline
        \TOOL{}         & \textbf{70.72}    & 74.94             & 79.62             & \textbf{76.34}    & \textbf{85.94}   \bigstrut[t] \\ % & 75.87 \\
        - code prompt   & 67.21             & 73.53             & 77.75             & 74.94             & 81.49             \\ % & * \\
        - single assert & 70.27             & \textbf{75.33}    & 79.39             & 76.01             & 80.85             \\ % & * \\
        - dyn. mutation & 68.85             & 72.13             & 78.68             & 75.35             & 81.84             \\ % & * \\
        % - dyn. pruning  & \textbf{70.49}    & 76.11             & 78.92             & 83.14             & \textbf{85.48}    \\ % & * \\
        - test ranking  & 63.23             & 65.80             & 72.36             & 64.87             & 71.89             \\ % & * \\
        - code ranking  & 69.32             & 75.17             & \textbf{81.96}    & 76.11             & 83.84 \bigstrut[b]\\ % & * \\
        \hline
    \end{tabular}
    }
    
\end{table}

\begin{table}[t]
    \centering
    \caption{Results of the HumanEval ablation studies. Metric \textit{p@k@m} stands for \textit{pass@k@m}.}
    \label{tab:rq3_human_eval}
    \resizebox{\linewidth}{!}{
    \begin{tabular}{l||c|c|c||c|c}
    \hline
        {\textbf{Tool}} & \emph{\textbf{p@1@1}} & \emph{\textbf{p@2@1}} & \emph{\textbf{p@5@1}} & \emph{\textbf{p@1@2}} & \emph{\textbf{p@1@5}} \bigstrut\\ % & \emph{pass@1@*} \\
        \hline
        \TOOL{}         & 55.27             & 64.59             & 73.29             & 68.94             & 84.47   \bigstrut[t] \\ % & 75.87 \\
        - code prompt   & \textbf{59.62}    & \textbf{65.21}    & 73.91             & 69.56    & 83.22             \\ % & * \\
        - single assert & 59.37             & 65.00             & 71.25             & 68.13             & \textbf{85.63}             \\ % & * \\
        - dyn. mutation & 57.76             & \textbf{65.21}    & \textbf{75.77}    & 66.45             & 80.00             \\ % & * \\
        % - dyn. pruning  & \textbf{70.49}    & 76.11             & 78.92             & 83.14             & \textbf{85.48}    \\ % & * \\
        - test ranking  & 48.44             & 55.90             & 65.83             & 49.68             & 57.76             \\ % & * \\
        - code ranking  & 56.52             & 63.97             & \textbf{75.77}    & \textbf{70.80}             & 82.60 \bigstrut[b]\\ % & * \\
        \hline
    \end{tabular}}

\end{table}

In order to examine the effects of various \TOOL{} components, we conduct ablation studies for each component individually.
Specifically, we evaluate the performance of \TOOL{} with \pass{k}{1} for $k \in \{1, 2, 5\}$,
as well as \pass{1}{m} for $m \in \{2, 5\}$.
We consider the following ablations:
\begin{itemize}[nosep]
    \item \TOOL{}: This is the default configuration as described in Section~\ref{setup}, 
    \item Code Prompt: ${\it TestGenPrompt} = choose(G)$, where we choose one of the suggested code suggestions in the prompt instead of $\texttt{pass}$. 
    \item Single Assert: ${\it StaticMutateTests} =$ none, where we do not perform any pruning of assertions. 
    \item Dynamic Mutation: ${\it DynMutateTests} =$ none.
    \item Test Ranking: ${\it TestRanking} =$ random.
    \item Code Ranking: ${\it CodeRanking} =$ none, where we use the standard (unranked) $\passk{k}$ metric over an unordered set of code suggestions. 
\end{itemize}

Table~\ref{tab:rq3_mbpp} and Table~\ref{tab:rq3_human_eval} present the result of ablation for the two benchmarks. 
For each metric, the cell corresponding to the highest-performing configuration is marked in bold.
Each configuration contributes differently to the evaluation, as we can see from the different metrics.
The default configuration was chosen to be the configuration that performed best on the \passdefault{} metric for the \emph{MBPP} dataset.
We prioritize the \passdefault{} metric since it is perhaps the most natural metric for code generation in an interactive setting since it relies on at most one user interaction and produces only one code suggestion as the final output.
Note that the default configuration also performs best on \passat{1}{2} and \passat{1}{5} for \emph{MBPP}. 
We observer, however, that for the \emph{HumanEval} dataset a different code prompt than \texttt{pass} yields higher numbers, performing the best for \passat{1}{1}, and \passat{2}{1}. 
Finding the optimal configuration that performs best on both benchmarks is subject of future work.

Although we do not find a configuration that performs well for all the metrics across the two different benchmarks, it is clear that most of the components have a non-trivial impact on the performance overall metrics.
Presenting the user with randomly picked tests from the set of test suggestions, rather than the top-ranked test, performs uniformly worse than the default configuration for both datasets. 
This indicates the importance of the test-ranking policy as a component.
Similarly, another component that is helpful for the default configuration is dynamic test mutation; removing it results in a dip in performance as shown in \autoref{tab:rq3_mbpp} for \emph{MBPP}. 
However, for HumanEval in \autoref{tab:rq3_human_eval} \pass{5}{1} is higher without dynamically mutated tests. This could be the case that the generation of many tests can adversely impact the rule-based test ranking.

Code ranking is clearly useful in the default configuration for the \passat{1}{m} metrics for \emph{MBPP} but performs worse for the
\passat{k}{1} metrics for $k \geq 2$. For \emph{HumanEval}, we observe again that code ranking can be important for 1 code suggestion, \emph{i.e.} high \passat{1}{1} and \passat{1}{2} with a non-\texttt{pass} prompt, but can have adverse effects for more.

The ablations also demonstrate that other components have a non-trivial effect on the evaluation metrics.
For example, disabling the static test mutation heuristic improves \pass{2}{1}, 
while using the code suggestions in the test generation prompt improves performance on \pass{1}{2}.
However, these configurations were not chosen to be the default since they all perform worse on the \passdefault{} metric.

\section{Threats}
\emph{Generalization of findings.}
We evaluate \TOOL{} using two popular and state-of-the-art research Python benchmarks for code generation tasks: MBPP and HumanEval. While both benchmarks exercise common programming patterns, they may not be representative of real-world software development. Our findings may not generalize to a different set of programs across different languages and problem domains.

\emph{Stability of model output.} As we have used OpenAI API to access the Codex model, we cannot control the stochasticity of the output by the model, and the model endpoints themselves are often updated or even discontinued. This poses a threat to the replicability of our study. We aim to mitigate this by releasing model generated output in the near future used in this study to improve reproducibility of our results. 

\emph{Interaction simulation.} 
To enable a large scale evaluation of the potential of \TOOL{} to improve correctness of generated code, we simulate user interaction by using the oracle of reference code implementations as common in interactive synthesis literature. 
However, our automated evaluation assumes the user is able to answer the generated tests, and cannot account for the cognitive effort of users undertaking the coding tasks. We plan to conduct a user study to evaluate such metrics in practice similar to prior works in PBE~\cite{zhang2020interactive}.

\section{Related work}
\label{sec:related}

AlphaCode~\cite{alphacode_2022} and CodeT~\cite{codet_2022} improve the \passk{k} metric by generating tests using LLMs (AlphaCode trains a new test-generation model) and then groups code suggestions by the set of tests they satisfy.
When suggesting code suggestions, only a single suggestion from each group is reported.
CodeT~\cite{codet_2022} refines the approach by scoring tests and code suggestions simultaneously by prioritizing tests that satisfy many code suggestions and prioritizing codes that satisfy many tests. 
Unlike \tappProblem{}, these approaches still \passk{k} metric and do not account for user interaction or provide any guarantees on suggested code. 
On the other hand, our test and code ranking components can benefit from the algorithms in CodeT --- we leave it as future work.

Scalable test generation for software has a rich history, and comprehensive coverage is outside the scope of this work. 
The dominating approaches for real-world code are based on variants of {\it feedback-driven random testing}~\cite{randoop-icse07} or on genetic programming~\cite{evosuite_2011}. These approaches are optimized for maximizing code coverage and finding runtime crashes.
However, these approaches are not directly applicable in \tappProblem{} scenario for two primary reasons (a) we do not start with an implementation of the method under test, and (b) it is critical to generate test oracles (or expected output) without access to the method definition. 
Neural approaches have shown promise recently in either synthesizing test oracles~\cite{atlas_2022, dinella2022toga} or high-coverage tests~\cite{lemieux2023codamosa} or generating an entire test~\cite{athenatest_2020},
We expect to harness these approaches to generate the {\it seed tests} that can be further mutated and ranked using suitable extensions to the algorithms presented in this work.

Finally, work on program synthesis~\cite{gulwani_2017, solar_lezama_2009} generates code that satisfies a formal specification either expressed as a logical specification or input-output examples~\cite{gulwani2011automating}.
Unlike program synthesis, LLMs generate code from informal specifications (our setup) and evaluate it through hidden tests or specifications. 
However, it would be interesting for future work to leverage user-provided tests to improve the quality of code generation, as explored in recent works~\cite{jain2022jigsaw, prose_multimodal_2021}.

Our work is closest to prior works in oracle-guided inductive synthesis~\cite{jha-icse-10,jha-acta-17} and interactive program synthesis~\cite{le2017interactive, interactive_synth_2020} in querying an oracle (reference implementation or users) for {\it distinguishing examples} in addition to initial set.
However, our work differs in several respects. First, most prior works appeal to an automatic symbolic engine (such as a constraint solver~\cite{interactive_synth_2020} or automata construction~\cite{zhang2020interactive}) to generate distinguishing example inputs for a pair of programs, which is inconceivable for general purpose imperative programming languages such as Python. 
Second, unlike classical uses of OGIS~\cite{jha-icse-10}, we only require the user to provide a 3-valued output instead of asking for output for a given input. This requires less effort on the part of a user, making our queries closer in spirit to membership queries in Angluin's  classical automata learning algorithm~\cite{angluin-87}.
Finally, unlike most program synthesis approaches that start with a handful of examples to express the intent, our formulation only starts with a natural language intent. 
Therefore, the augmented test cases are not only valueable to arrive at the correct code, but also serve as regression unit tests for maintaining the code. 
%\section{Future Work}
%\label{sec:future}

\section{Conclusions}
\label{sec:concl}

In this work, we study the workflow of test-driven interactive code generation using LLMs. 
We explore various dynamic approaches for improving the effectiveness of a purely baseline LLM-based solution, and illustrate non-trivial improvements on code and test generation accuracy with different components for mutating, pruning and ranking test and code suggestions.

In future works, we plan to collect and evaluate our approach on real-world benchmarks such as CoderEval~\cite{yu2023codereval} and nl2fix~\cite{fakhoury2023towards}. 
We also leave as future work a \emph{user study} of \TOOL{} to complement our findings from the simulated quantitative evaluation.
Finally, we plan to explore how our approach can be extended to richer forms of formal specifications beyond tests (e.g., parameterized tests, and procedure summaries) given a scalable generator and checker of such forms of specifications.

%%%%%%%%%%%%%%%%%%%%%%%%%%%%%%%%%%%%

\bibliographystyle{ieeetr}
\bibliography{refs}

\end{document}